\documentclass[conference]{IEEEtran}
\IEEEoverridecommandlockouts
\usepackage{cite}
\usepackage{amsmath,amssymb,amsfonts}
\usepackage{algorithmic}
\usepackage{graphicx}
\usepackage{siunitx}
\usepackage{textcomp}
\usepackage{mathtools}
\usepackage{xcolor}
\usepackage[all, warning]{onlyamsmath}
\usepackage{color, graphicx}

\newcommand{\myboldmath}[1]{\boldsymbol{#1}}
\newcommand{\bx}{\myboldmath{x}}

\usepackage[whole]{bxcjkjatype} % Japanese
\def\BibTeX{{\rm B\kern-.05em{\sc i\kern-.025em b}\kern-.08em
    T\kern-.1667em\lower.7ex\hbox{E}\kern-.125emX}}

\title{Three-dimensional hand guidance \\ by midair haptic display}

\author{Koya Hiura$^{1}$, Shun Suzuki$^{1}$, Tao Morisaki$^{1}$, Masahiro Fujiwara$^{1}$, Yasutoshi Makino$^{1}$ and Hiroyuki Shinoda$^{1}$% <-this % stops a space

\thanks{$^{1}$All members are with Graduate School of Frontier Sciences,
        The University of Tokyo, 5-1-5 Kashiwa-shi, Chiba, 277-8561, Japan
        {\tt\small hiura@hapis.k.u-tokyo.ac.jp},
        {\tt\small suzuki@hapis.k.u-tokyo.ac.jp},
        {\tt\small morisaki@hapis.k.u-tokyo.ac.jp},
        {\tt\small Masahiro\_Fujiwara@hapis.k.u-tokyo.ac.jp}, 
        {\tt\small yasutoshi\_makino@hapis.k.u-tokyo.ac.jp}, 
        {\tt\small hiroyuki\_shinoda@hapis.k.u-tokyo.ac.jp}
        }%
}

\begin{document}

\maketitle

\begin{abstract}

Guiding human movements using tactile information is one of the promising applications of haptics. Using midair ultrasonic haptic stimulation, it is possible to guide a hand without visual information. However, the information of movement shown by conventional methods was partial. It has not been shown a method to guide a hand to an arbitrary point in three-dimensional space. In this study, we propose a method of guiding the hand to the top of a virtual cone presented haptically and evaluate the effectiveness of the method through experiments. As a result, the method guided the participant's hand to the goal in a 30\,cm-cube workspace with an error of 64.34\,mm.

\end{abstract}


\section{Introduction}

\begin{figure}[tbp]
    \centering
    \includegraphics[width=\linewidth]{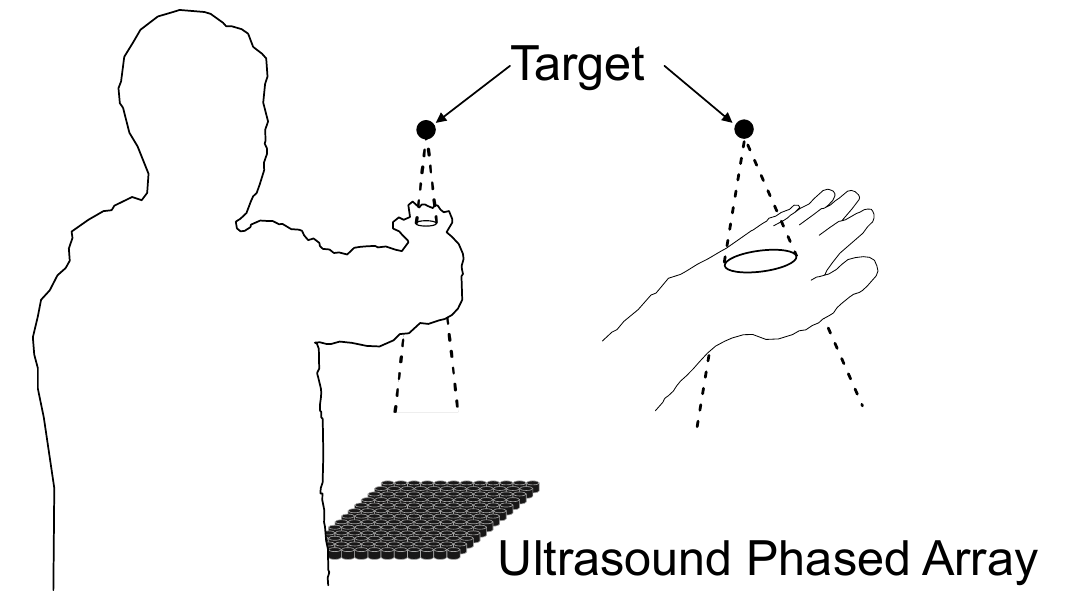}
    \caption{Motion induction by a haptic cone}
    \label{fig:concept}
\end{figure}

This study proposes a method for non-contact hand guidance to an arbitrary point in space using an ultrasound phased array.
Ultrasound phased array focus ultrasound in a space about the size of the wavelength and can present tactile sensations by the radiation pressure\cite{iwamoto2008non, hoshi2010noncontact}.
We propose a method of user hand guidance using a haptic cone based on this airborne ultrasound tactile perception, as shown in Fig~\ref{fig:concept}.
Users can move their hands toward the apex while feeling the change in the cross-sectional area on the palm.

Several methods have already been proposed for non-contact guidance of the hand using midair haptic stimulation.
However, no method has been shown to guide a hand to an arbitrary point in three-dimensional space.
Yoshimoto et al. showed that the palm of a human hand could follow the movement of a tactile stimulus point~\cite{yoshimoto2019midair}.
They also proposed a method of hand guidance using this human faculty; they guided the user's hand by having the user follow the focal point as it moved toward the destination.
Suzuki et al. generated tactile stimulus lines in space using focused ultrasound and used them as handrails for guidance~\cite{suzuki2019midair}.
The guidance in these previous studies used stimuli in the direction parallel to the palm, where spatial patterns are easily perceived.
Therefore, the guidance is limited to a two-dimensional plane.
A previous study~\cite{suzuki2018haptic} showed that the hand could be guided perpendicular to the palm using an ultrasound bessel beam.
However, because this study used a one-dimensional line, they could not tell the user the direction in which to move.
Freeman et al. propose the HaptiGlow system, which combines ultrasonic tactile and visual feedback~\cite{freeman2019haptiglow}.
This system presents the user with a tactile sweet spot where gesture recognition is easy, and haptic feedback is effectively provided. 
This system also uses a conical tactile pattern.
However, the sweet spot is the base of the cone, where the tactile sensation is strong over a wide area.
It is difficult to accurately understand the size of the bottom area with a palm, hence the definition of the target position might be ambiguous.
In addition, the target positions were placed within a workspace of \SI{6}{cm} in the $z$-direction and \SI{8}{cm} in the $xy$-direction.
The guidance performance using a haptic cone in a larger workspace is still unknown.

In this study, we present a circular pattern stimulus that shows the cross-section of a cone and a hand, similar to~\cite{freeman2019haptiglow}.
We propose a macroscopic guidance method that guides the user's hand to an arbitrary point in the three-dimensional space.
However, the purpose is to guide the user to a specific point in the three-dimensional space. 
Therefore, contrary to a previous study~\cite{freeman2019haptiglow}, we propose a method in which the apex of the cone is the target point.
In addition, we evaluate the guidance performance in a workspace of \SI{30}{cm} cube.
The proposed method can inform the user of a specific location in the space in situations where vision is unavailable.
It can also assist the user in locating the reference point for gesture recognition by limiting the workspace of the 3D interface.
This can be applied to improve the accuracy of measuring the user's movements and facilitate the interpretation of the user's intentions.

\section{Related Works}

Many studies have been conducted on human behavioral guidance using tactile sensation.
Tactile information is as effective as auditory and visual information in moving the body to a specific position.
Tactile information has the advantage of being less affected by the external environment, such as noise and sunlight, and is more intuitive.
Braille blocks and handrails are typical examples of inducing human movement using tactile information.
In these cases, the user can move to a predetermined destination by following a guide built into the environment.

Some studies have been conducted on wearable or portable devices that allow users to set destinations programmatically.
Cosgun et al. developed a tactile belt worn on the waist~\cite{cosgun2014guidance}.
This belt tells the user rotation and movement information by vibration patterns from transducers on the belt.
Baldi et al. have developed a tactile armband that vibrates to give three directions to a walking human, turn left, turn right, and go straight, and have demonstrated its effectiveness~\cite{baldi2017haptic}.
There are also studies using tactile stimuli on the palm.
Amemiya et al. developed a handheld device that creates the sensation of being pulled by asymmetrically vibrating a weight~\cite{amemiya2008lead}.
This device enables intuitive guidance by providing the sensation of being pulled toward the destination.

\begin{figure}[tbp]
    \centering
    \includegraphics[width=0.8\linewidth]{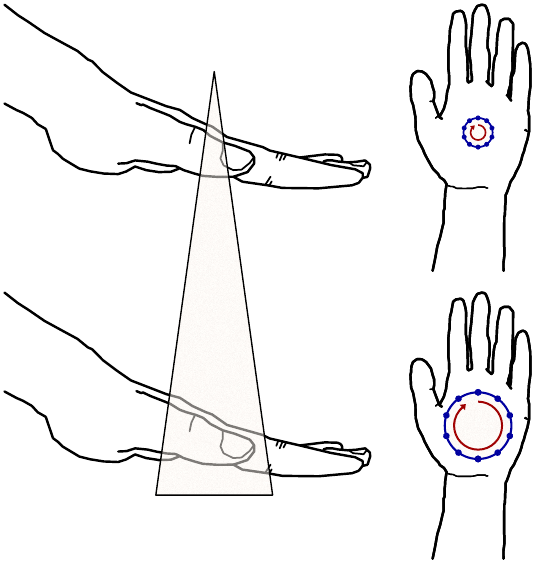}
    \caption{Proposed method: Make the user perceive a cone by presenting a circle with a radius corresponding to the hand's position.}
    \label{fig:method}
\end{figure}

In addition, as mentioned in the introduction, a method of guiding a user's movement without wearing or holding a device has been proposed by using ultrasound in the air~\cite{yoshimoto2019midair,suzuki2018haptic,suzuki2019midair,freeman2019haptiglow}. 
These techniques do not require the user to have a device.
Therefore, it is a convenient method for the general public people in public places.

\section{Method}

In this study, we propose a method to guide the hand to the apex of a virtual haptic cone.
As shown in Fig.~\ref{fig:method}, when a user touches this cone, the user perceives its cross-section.
The only information the user needs to know in advance is that the apex is the taget point.
Therefore, the user can reach the destination by moving the hand in the direction where the perceived cross-sectional area becomes smaller.
First, at the starting point, the user moves the hand around a little to find the direction in which the radius becomes smaller.
The user then moves the hand toward the apex while trying to feel the circle around the center of the palm.
This method is based on the image of a cone; thus, intuitive guidance can be achieved.

\begin{figure}[tbp]
  \centering
  \includegraphics[width=0.7\linewidth]{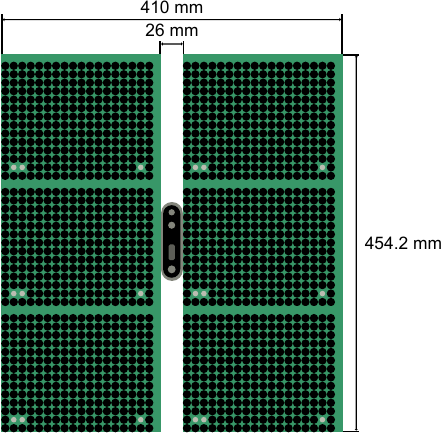}
  \caption{Experiment setup. A $\SI{410}{mm} \times \SI{454.2}{mm}$ phased array was configured from six AUTD3 units~\cite{suzuki2021autd3}. A depth sensor is placed in the center of the phased array to track the position of the user's hand.}
  \label{fig:exp_setup}
\end{figure}

In this study, focused ultrasound is generated using an ultrasonic phased array.
When a person touches focused ultrasonic waves, the acoustic radiation pressure causes a tactile sensation.
In the case of \SI{40}{kHz} ultrasound, people perceive a focus of about \SI{1}{cm}.
This paper adopts AUTD3, a scalable ultrasound phased array~\cite{suzuki2021autd3}.

To guide the movement, it is necessary to present the circle by ultrasonic tactile stimulation.
In this study, we used the Spatio-Temporal Modulation (STM) method~\cite{takahashi2018lateral, kappus2018spatiotemporal}.
It is a technique of moving a focal point with constant amplitude at an appropriate frequency on the skin.
The STM method is capable of clearly representing the shape.
In this study, $N$ points are sampled around the cross-section of a cone.
The focal point is moved by switching the sampling points with the frequency $f_s$ to provide a tactile sensation.
The cross-section is presented $f=f_s/N$ times per unit of time.

Determining $N$, $f$, and $f_s$ is important for best performance. 
The stimulus frequency $f$ should be a value that humans easily perceive, and the location of the stimulus should be easily identifiable.
In addition, we need to consider the influence of the distance between adjacent stimulus points on perception.
Frier et al. showed that large sampling numbers and circular frequency values don't necessarily improve tactile perception~\cite{frier2019sampling}. 
In addition, the ultrasound transducer is a resonator with a high $Q$-factor.
Therefore, a certain amount of time is needed to switch the focus of the ultrasonic phased array~\cite{suzuki2020reducing}.
When $f_s$ exceeds the limit speed that the transducer can follow, the ultrasonic amplitude becomes weak.

In this study, however, we don't discuss the problem of finding this optimal condition in detail. 
First, we set $N=10$ as $N$ sufficient to display the cross-section.
From previous studies~\cite{frier2019sampling}, it is known that when $N = 10$, the perceived intensity is highest at low frequencies below $f = \SI{10}{Hz}$.
On the other hand, the rendering frequency $f$ should be high so that users perceive the cones as quickly as possible.
Therefore, we chose $f = \SI{10}{Hz}$ in this experiment.
Thus, the sampling frequency is $f_s=fN = \SI{100}{Hz}$.
This is slow enough to switch the focal point ~\cite{suzuki2020reducing}.

\begin{figure}[tbp]
 \begin{minipage}[t]{0.33\columnwidth}
  \centering  
  \includegraphics[width=\linewidth]{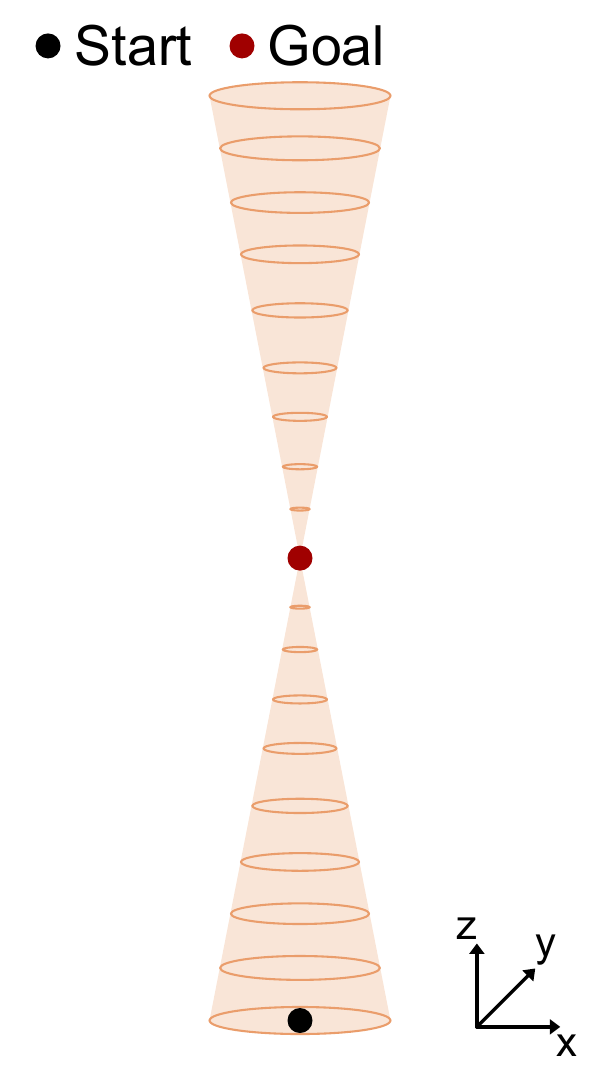}
  (a)
 \end{minipage}%
 \begin{minipage}[t]{0.33\columnwidth}
  \centering  
  \includegraphics[width=\linewidth]{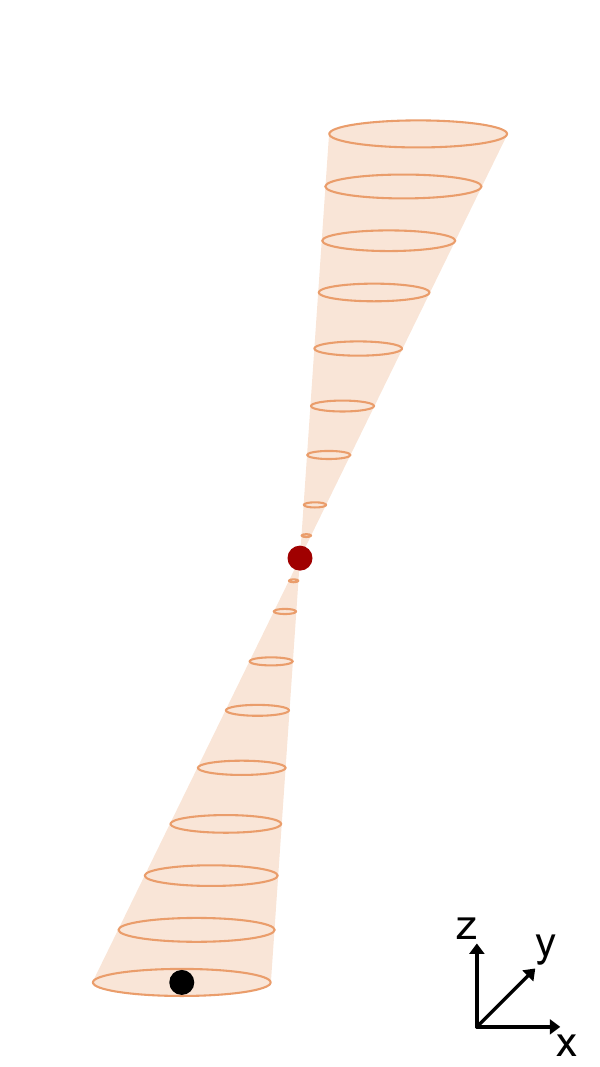}
  (b)
 \end{minipage}%
 \begin{minipage}[t]{0.33\columnwidth}
  \centering  
  \includegraphics[width=\linewidth]{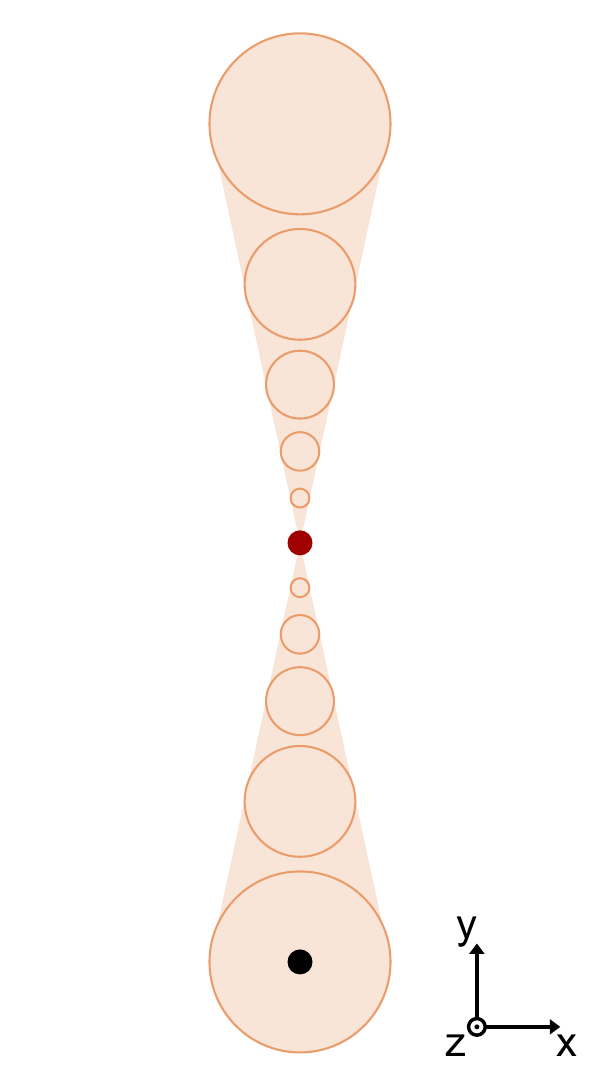}
  (c)
 \end{minipage}
 \caption{(a) a cone presented to the user. (b) an oblique cone is presented if the goal position is at an angle to the $z$-axis. (c) If both the start and goal are in the same $xy$ plane, a shape that an oblique cone is projected onto the $xy$ plane is presented.}
 \label{fig:exp_cone}
\end{figure}

\begin{figure}[tbp]
    \centering
    \includegraphics[width=0.3\linewidth]{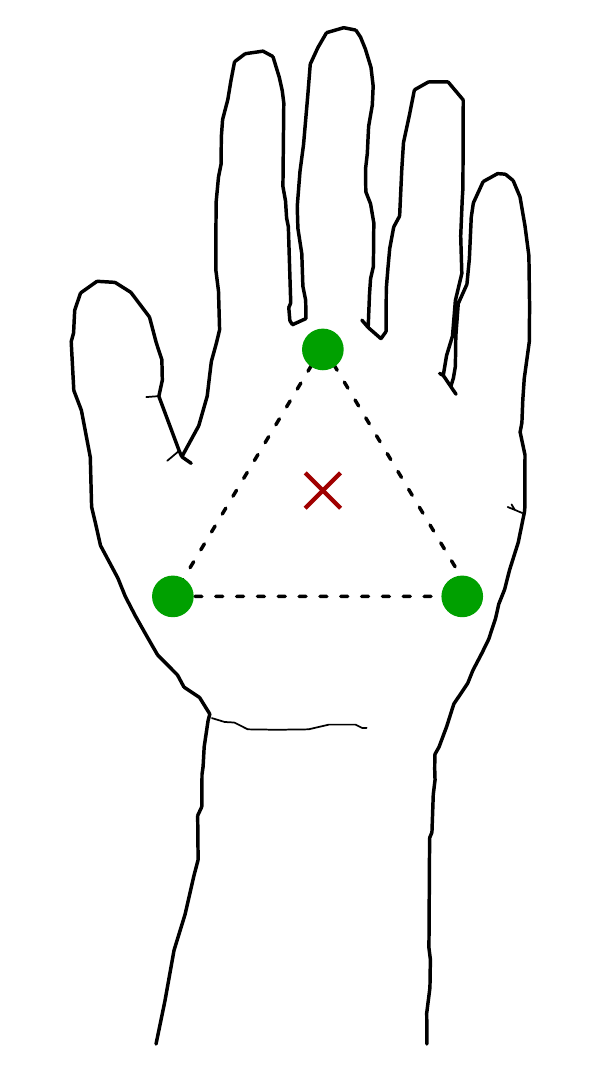}
    \caption{Handtracking method used in this paper. The center of gravity of the three colored stickers is defined as the position of the user's hand.}
    \label{fig:exp_handtracking}
\end{figure}

\section{Experiment}

\subsection{Setup}

Fig.~\ref{fig:exp_setup} shows a prototype system used in this paper.
We used six AUTD3 units~\cite{suzuki2021autd3}.
A depth sensor (RealSense Depth Camera D435, Intel Corporation) for sensing the user's hand is placed in the center of the phased array.
A coordinate system is adopted in which the $x, y$ axes are on the surface of the phased array and the $z$ axis is perpendicular to it.

As shown in Fig.~\ref{fig:exp_cone}(a), we present a haptic cone by rendering the circumference of the cross-section.
In this study, we assume the user's hands are parallel to the $xy$ plane.
Therefore, when the start and goal points are not aligned on the $z$ axis, we present an oblique cone instead of a tilted cone (Fig.~\ref{fig:exp_cone}(b)).
If the start and goal points are on the $xy$ plane, we present the shape of the oblique cone projected onto the $xy$ plane (Fig.~\ref{fig:exp_cone}(c)).

We define the used's hand as $\bx_u = (x_u,y_u,z_u)$, the starting point as $\bx_s = (x_s,y_s,z_s)$, and the goal point as $\bx_g = (x_g,y_g,z_g)$.
The formula for the circumference presented to the user is as follows;

\begin{align}
    \left\|(x,y) - (x_c, y_c)\right\|^2 = \left(kR\right)^2,
\end{align}
\begin{align}
    (x_c, y_c) &\coloneqq k(x_s, y_s) + (1-k)[(x_g, y_g) - (x_s, y_s)],\\
    k &\coloneqq \begin{cases}
            \frac{\|z_g - z_u\|}{\|z_g - z_s\|} & (z_s \neq z_g)\\
            \frac{\sqrt{(x_g - x_u)^2 + (y_g - y_u)^2}}{\sqrt{(x_g - x_s)^2 + (y_g - y_s)^2}} & (z_s = z_g)
        \end{cases},
\end{align}
where $(x_c, y_c)$ is the center of the circle to be presented.
The $z$-coordinate of the presented circle is the plane where the user's hand is located $z=z_u$ in the case of $z_s \neq z_g$ and $z=z_u=z_s$ in the case of $z_s = z_g$.
The parameter $k$ is $k=1$ at the start and $k=0$ at the goal.
Thus, the radius of the circle decreases as it approaches the goal.
However, if the radius of the circle becomes too small, the STM tactile sensation cannot be presented.
Therefore, we set the radius to be no smaller than \SI{1}{cm}.
In this study, the radius $R$ of the base of the cone is set to $R=\SI{30}{mm}$.
Note that in the case of $z_s \sim z_g$, the radius of the circle and its center are very sensitive to the user's hand motion.
Therefore, some practical implementation is needed, but we do not mention this problem in this paper.

To determine the user's hand position, we placed three stickers on the palm so that the center of gravity of the colored stickers was in the center of the palm, as shown in Fig.~\ref{fig:exp_handtracking}.
We used this center of gravity as the coordinates of the hand.

\subsection{Procedure}

\begin{figure}[tbp]
    \centering
    \includegraphics[width=0.6\linewidth]{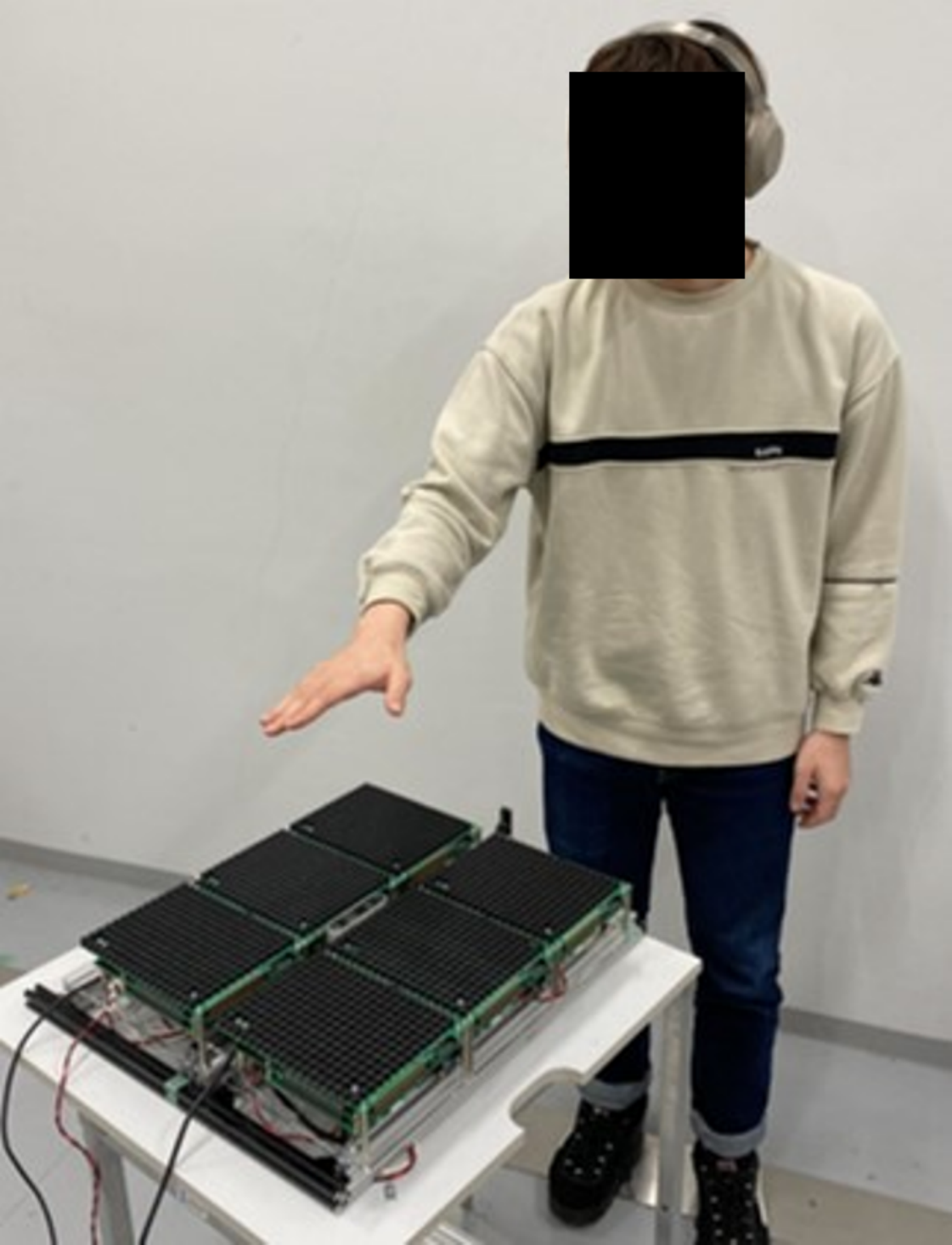}
    \caption{A view of the subject experiment}
    \label{fig:exp_human}
\end{figure}

\begin{figure}[tbp]
      \centering
        \includegraphics[width=0.8\linewidth]{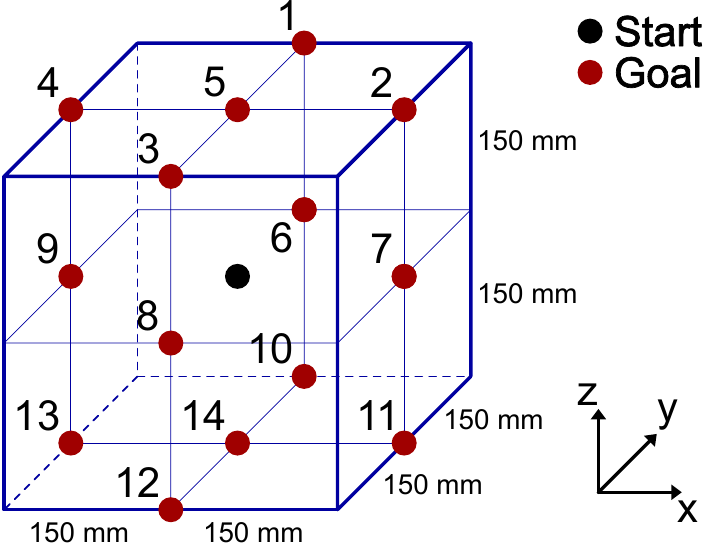}
        \caption{Fourteen guidance directions of the experiment}
        \label{fig:14direction}
\end{figure}

The participant stands in front of the system and has his dominant hand 400 mm above the phased array (Fig.~\ref{fig:exp_human}).
This point was used as the starting point for all experiments.
The participants wear headphones with white noise to eliminate any auditory cues.
The participants were instructed to move their hands toward the apex after feeling a tactile sensation and press the switch with their non-dominant hand when they reached the apex.
We measured the hand position during this trial and the time to reach the apex.
As shown in Fig.~\ref{fig:14direction}, we set the fourteen goals and present a haptic cone corresponding to the goal in random order.
We defined these fourteen trials as a set and conducted three sets on each participant.
If the participant couldn’t reach the goal within 30 seconds, we interrupted the trial, and started the next trial.
Ten men and three women with an average age of 24.7 participated in the experiment.
In addition, there was one left-handed person, and all others were right-handed.
They all had normal tactile abilities, and twelve participants had experienced midair haptics.

\subsection{Result}

Table~\ref{table:rate} shows the trial completion rate, the rate participants could complete the trial in 30 seconds.
The following results exclude trials for which the trial could not be completed.

We define the error in three-dimentional $\varepsilon_{xyz}$ as
\begin{align}
    \varepsilon_{xyz} = \|\bx_f - \bx_g\|,
\end{align}
where $\bx_f=(x_f, y_f, z_f)$ is the hand position where the participant judged that they had reached the goal.

Also, we difine the error in two-dimentional $\varepsilon_{xy}$ as
\begin{align}
    \varepsilon_{xy} = \|(x_f, y_f) - (x_c, y_c)\|.
\end{align}

\begin{figure}[t]
 \begin{minipage}[t]{0.33\columnwidth}
  \centering
  \includegraphics[width=\linewidth]{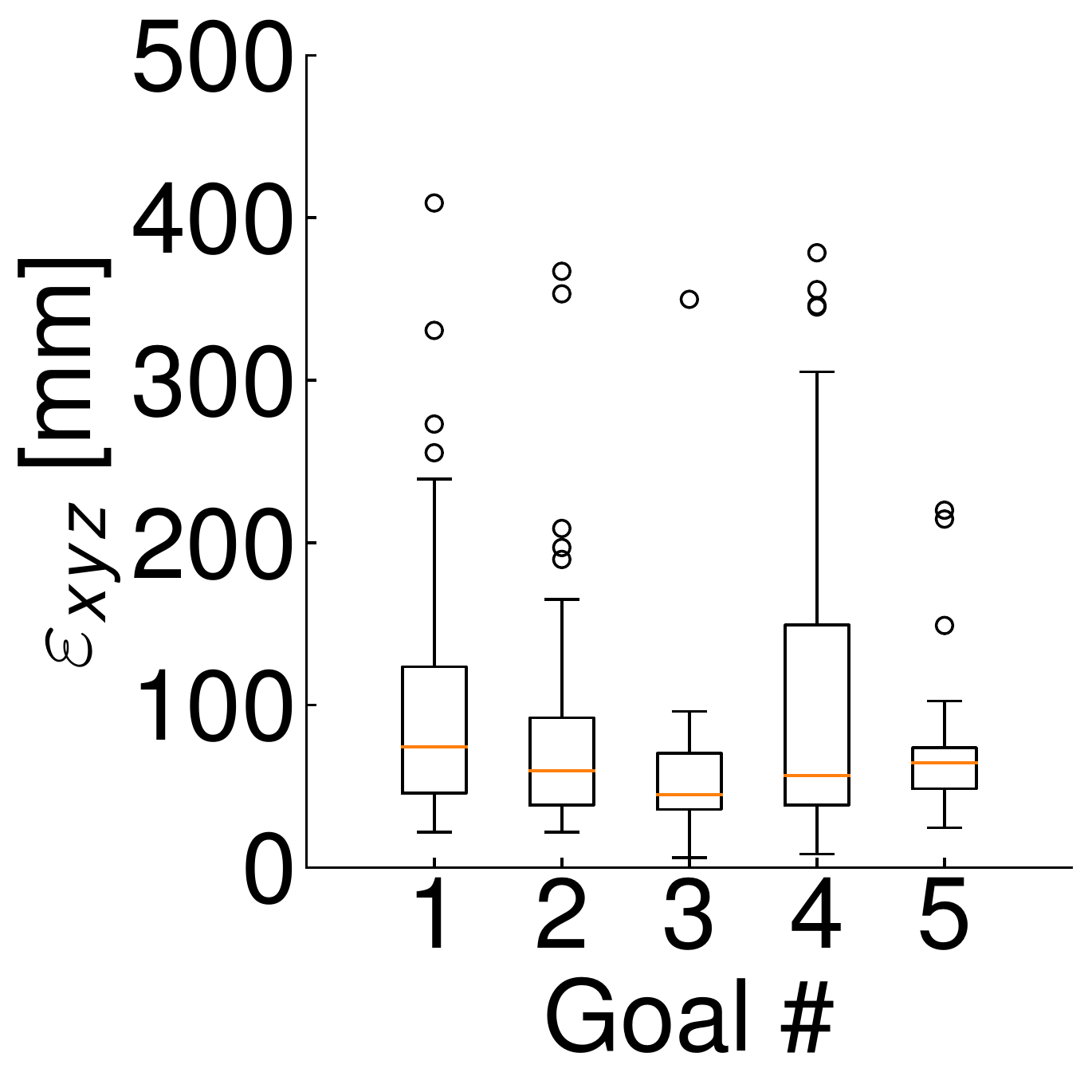}
 \end{minipage}%
 \begin{minipage}[t]{0.33\columnwidth}
  \centering
  \includegraphics[width=\linewidth]{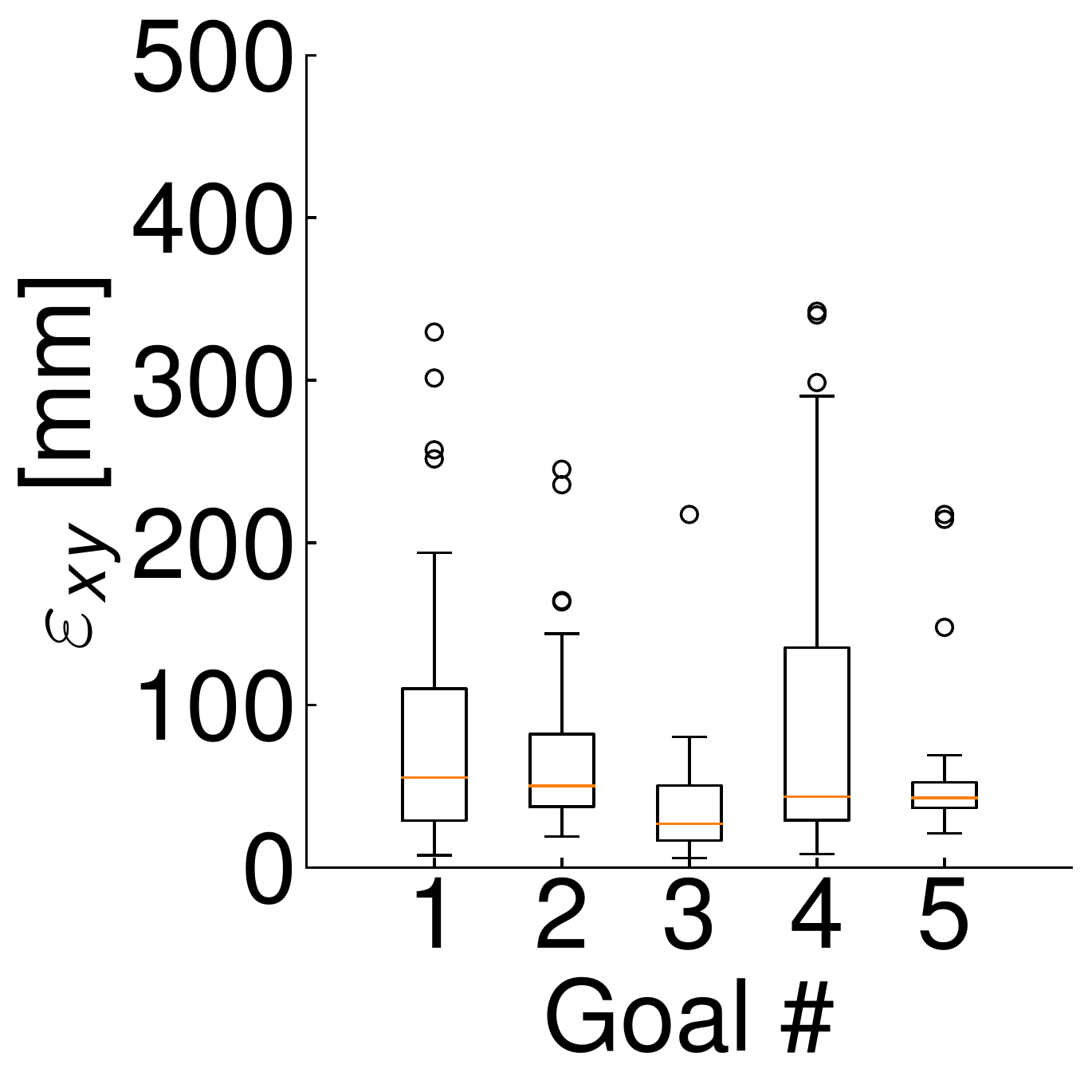}
 \end{minipage}%
 \begin{minipage}[t]{0.33\columnwidth}
  \centering
  \includegraphics[width=\linewidth]{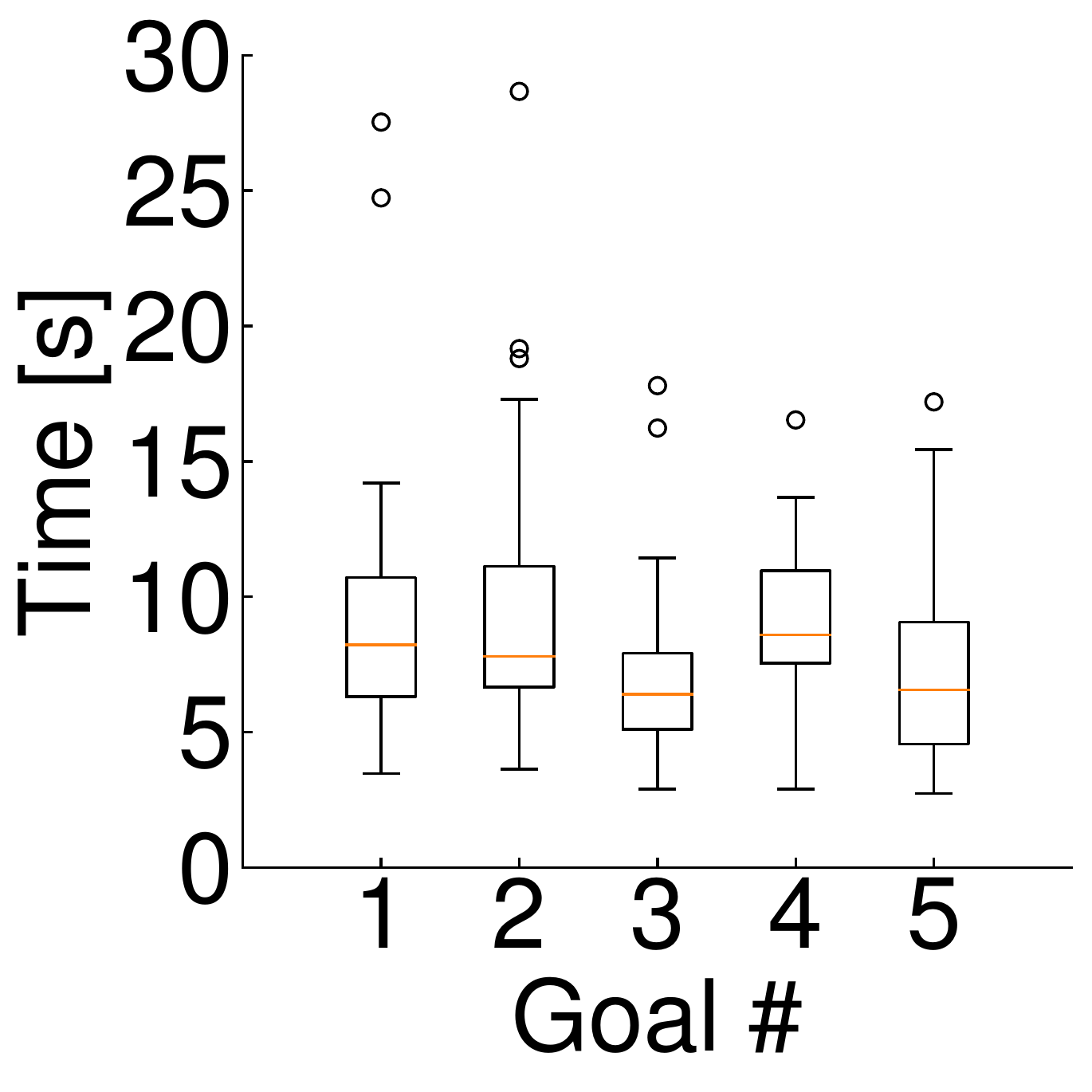}
 \end{minipage}\\
 \begin{minipage}[t]{0.33\columnwidth}
  \centering
  \includegraphics[width=\linewidth]{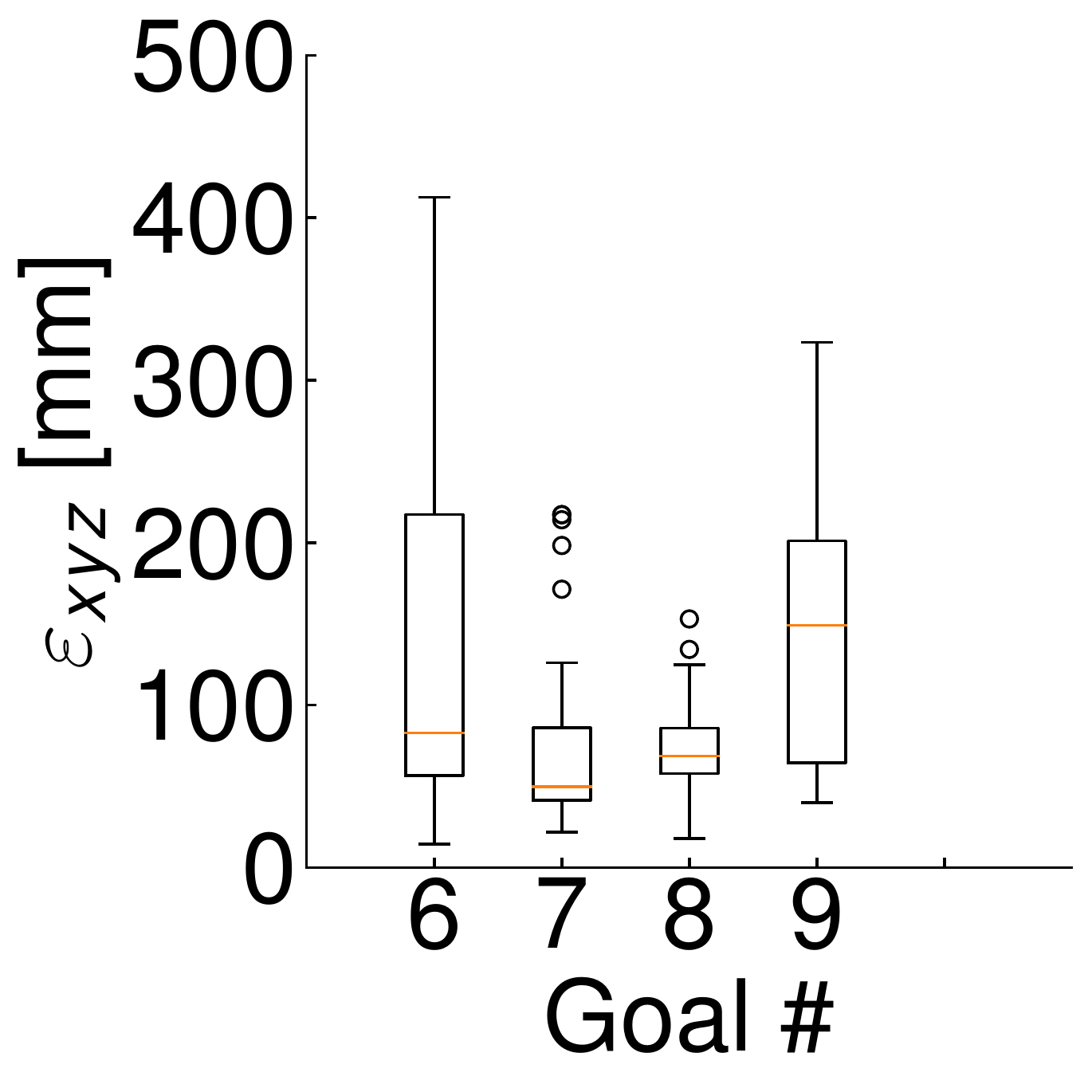}
 \end{minipage}%
 \begin{minipage}[t]{0.33\columnwidth}
  \centering
  \includegraphics[width=\linewidth]{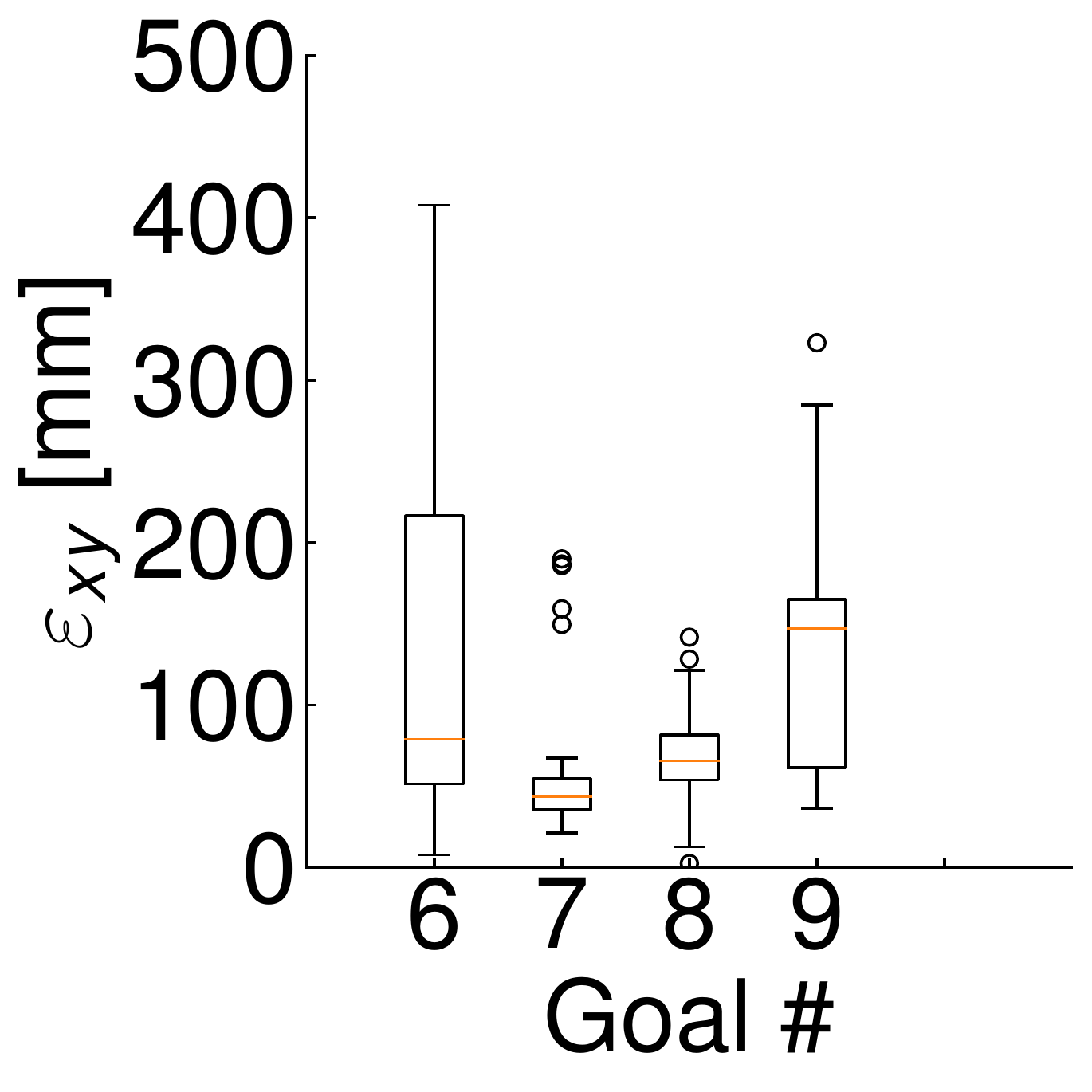}
 \end{minipage}%
 \begin{minipage}[t]{0.33\columnwidth}
  \centering
  \includegraphics[width=\linewidth]{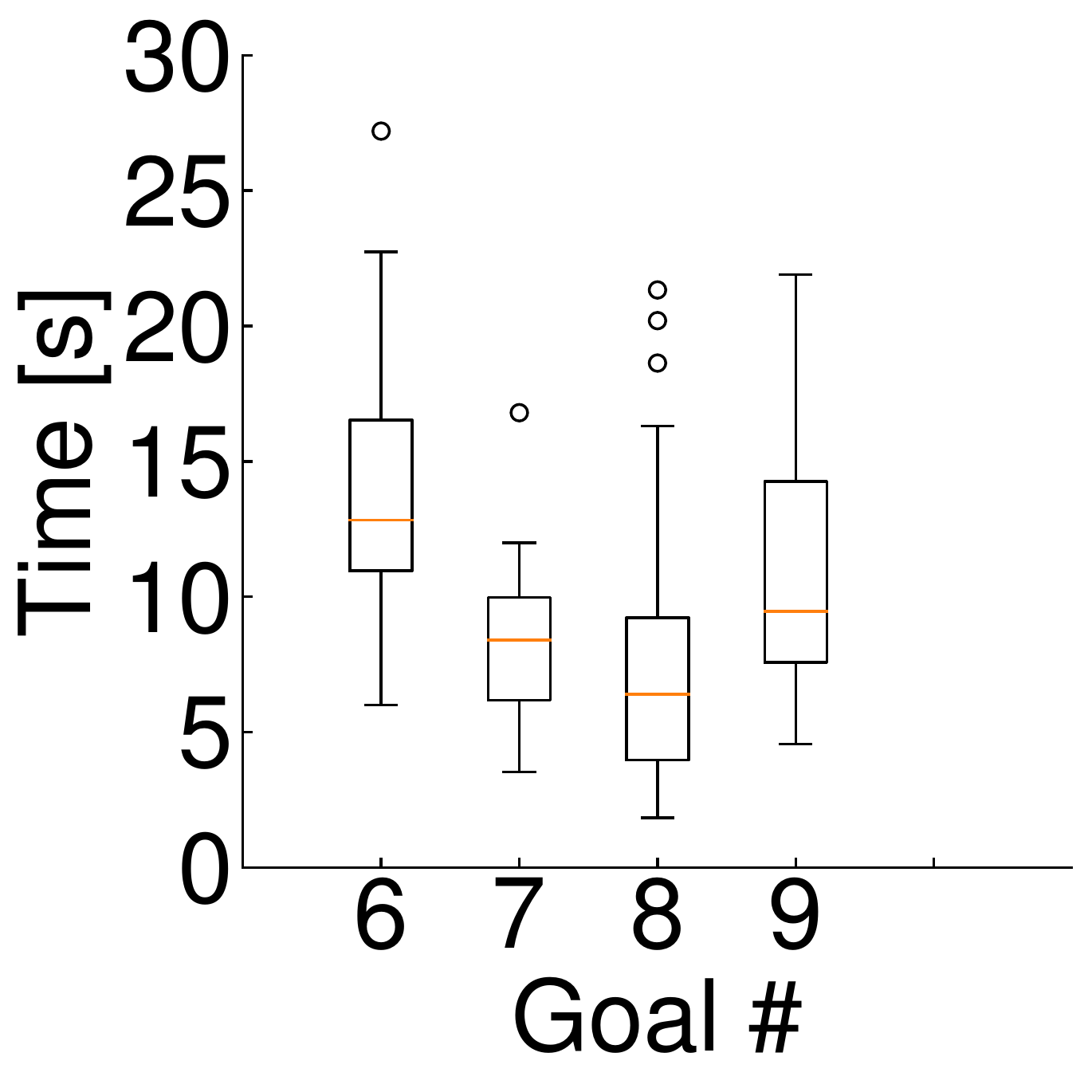}
 \end{minipage}\\
 \begin{minipage}[t]{0.33\columnwidth}
  \centering
  \includegraphics[width=\linewidth]{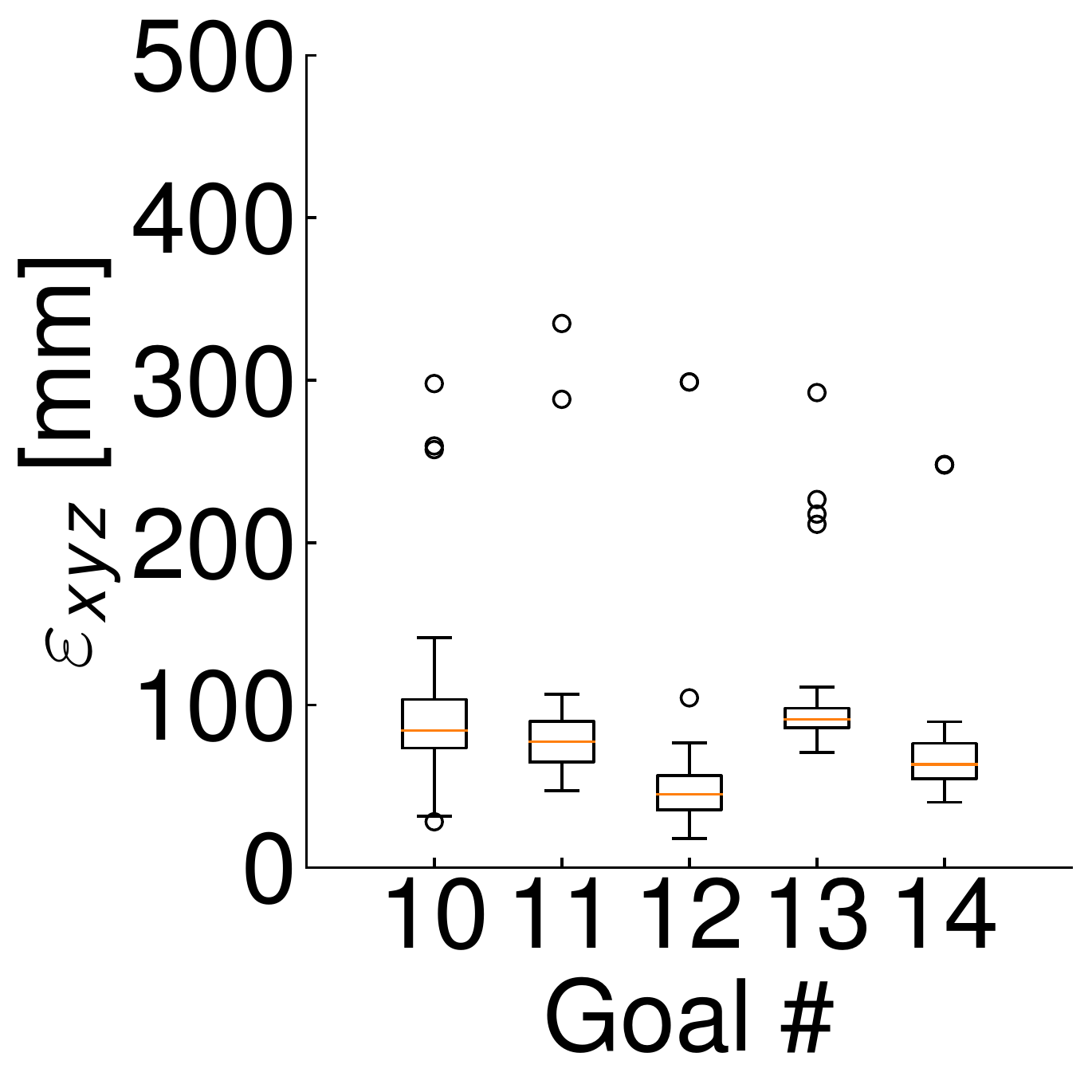}
 \end{minipage}%
 \begin{minipage}[t]{0.33\columnwidth}
  \centering
  \includegraphics[width=\linewidth]{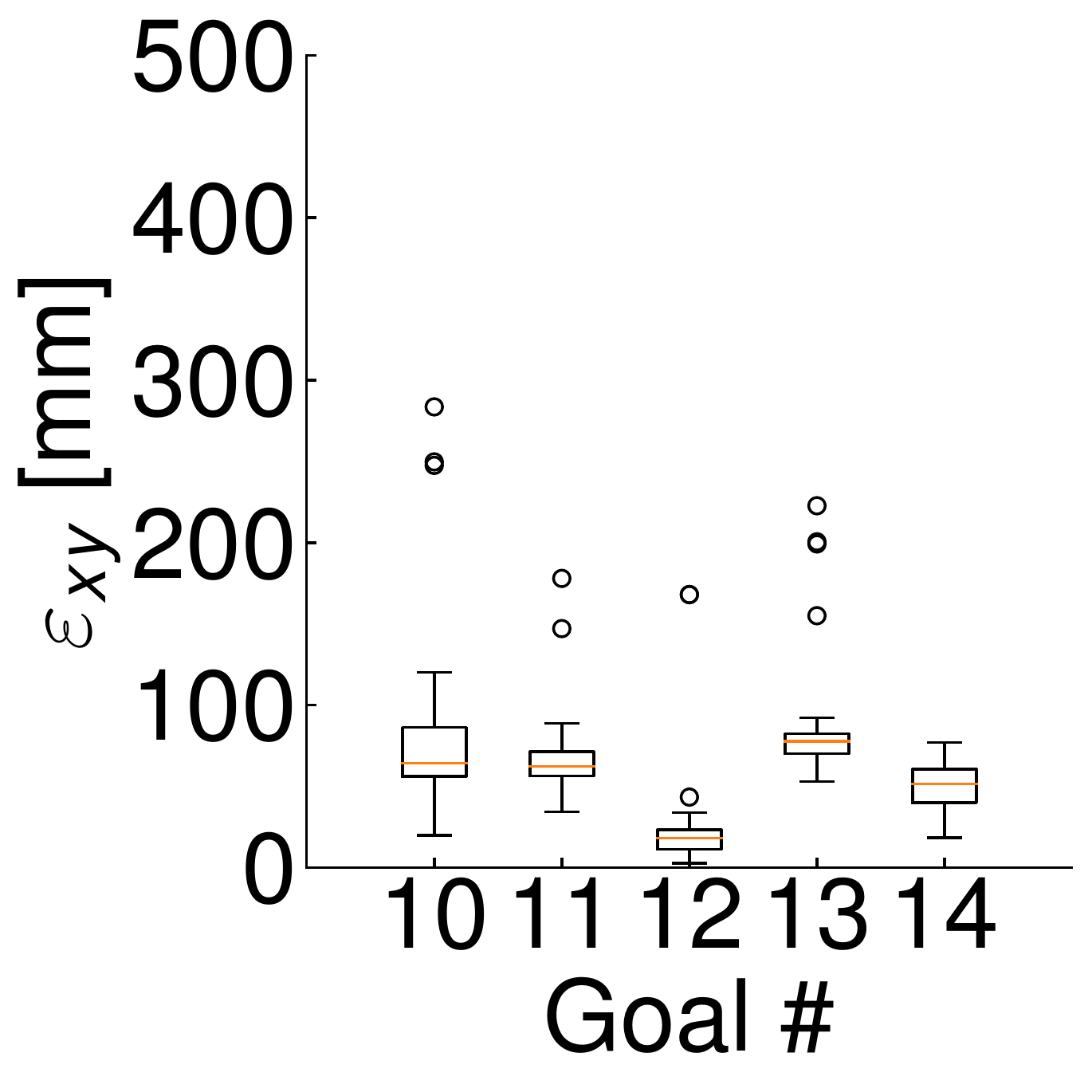}
 \end{minipage}%
 \begin{minipage}[t]{0.33\columnwidth}
  \centering
  \includegraphics[width=\linewidth]{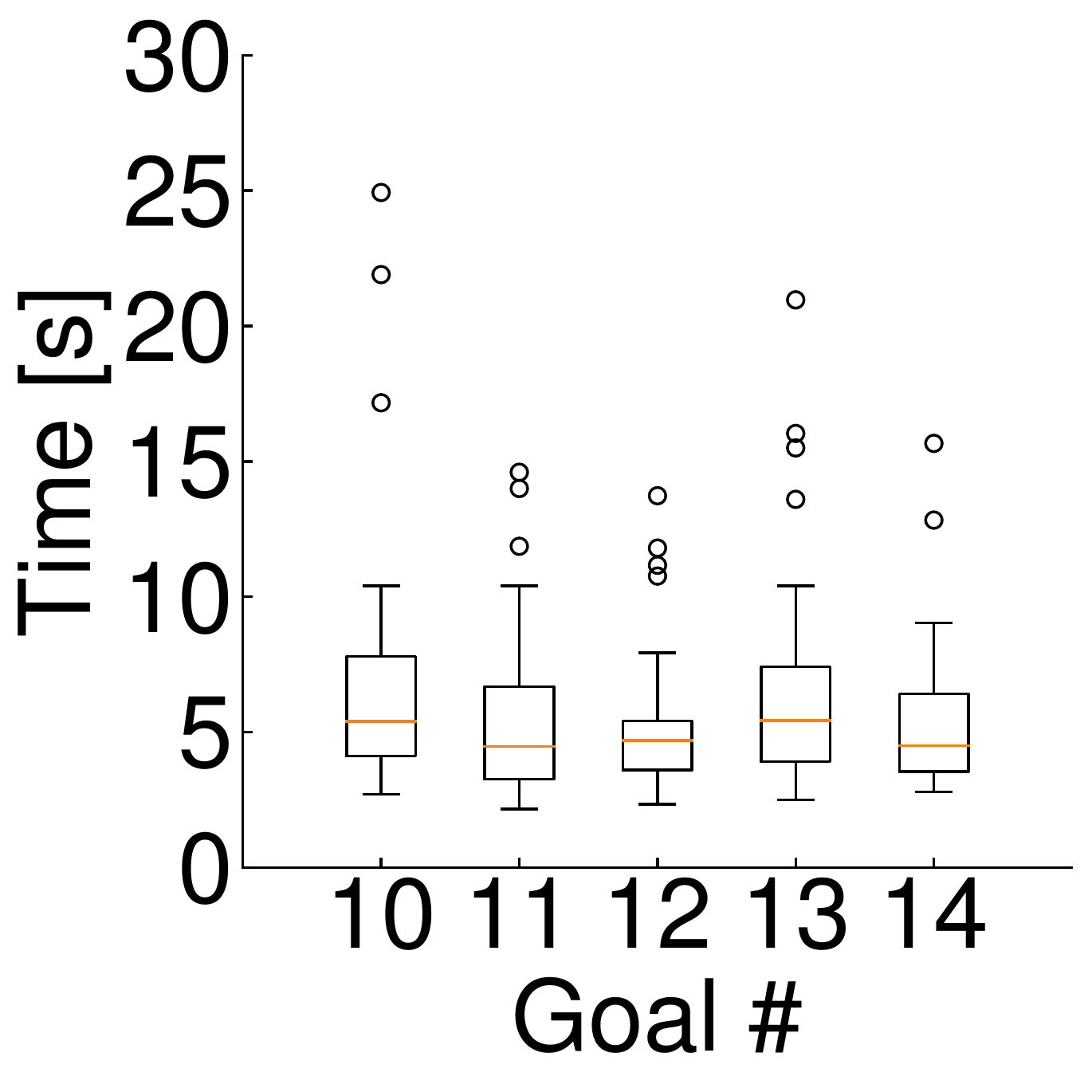}
 \end{minipage}%
 \caption{The deviations from the goal and elapsed time to finish trials}
 \label{fig:eps_xyz_and_time}
\end{figure}

\begin{figure}[t]
 \begin{minipage}[t]{0.5\columnwidth}
  \centering
  \includegraphics[width=\linewidth]{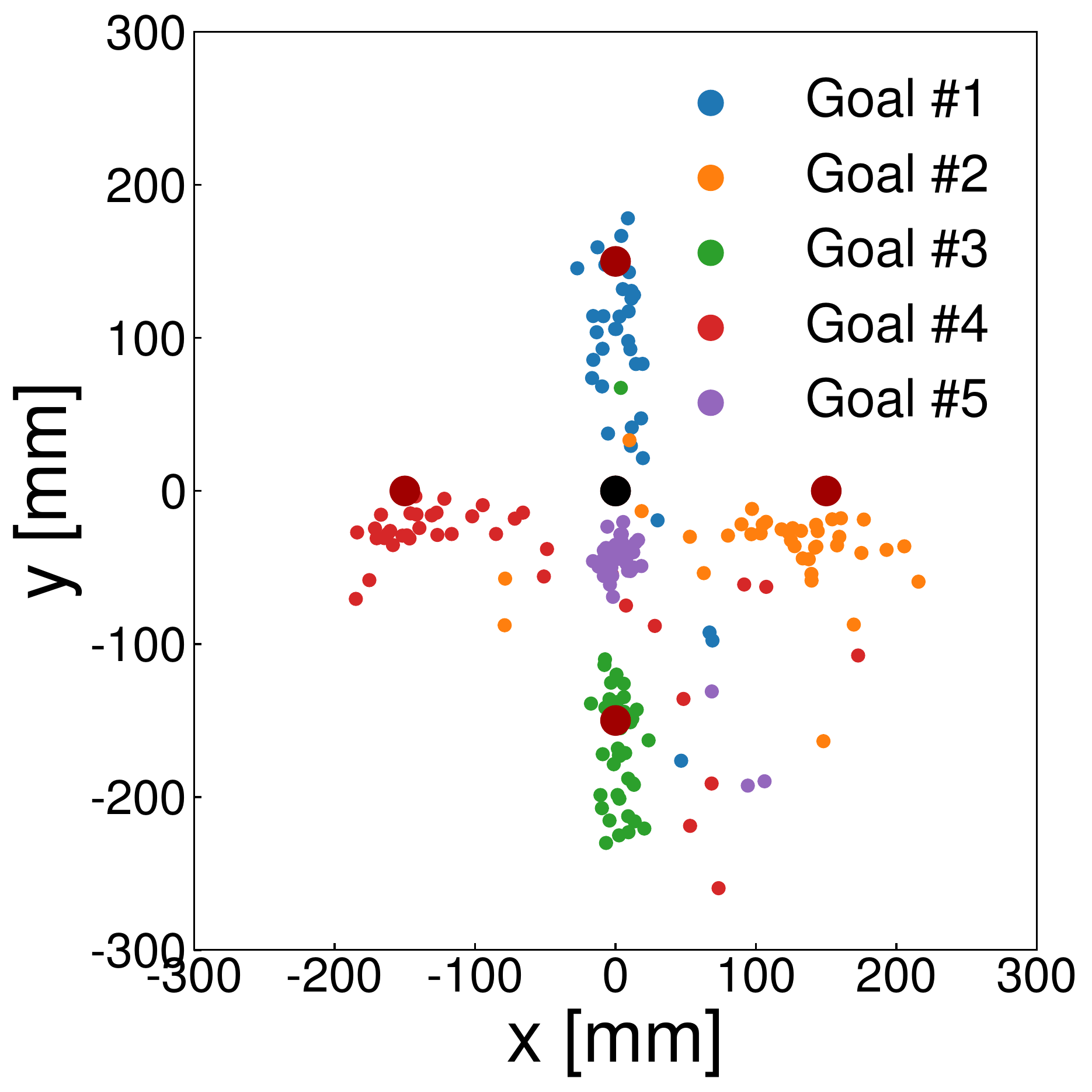}
 \end{minipage}%
 \begin{minipage}[t]{0.5\columnwidth}
  \centering
  \includegraphics[width=\linewidth]{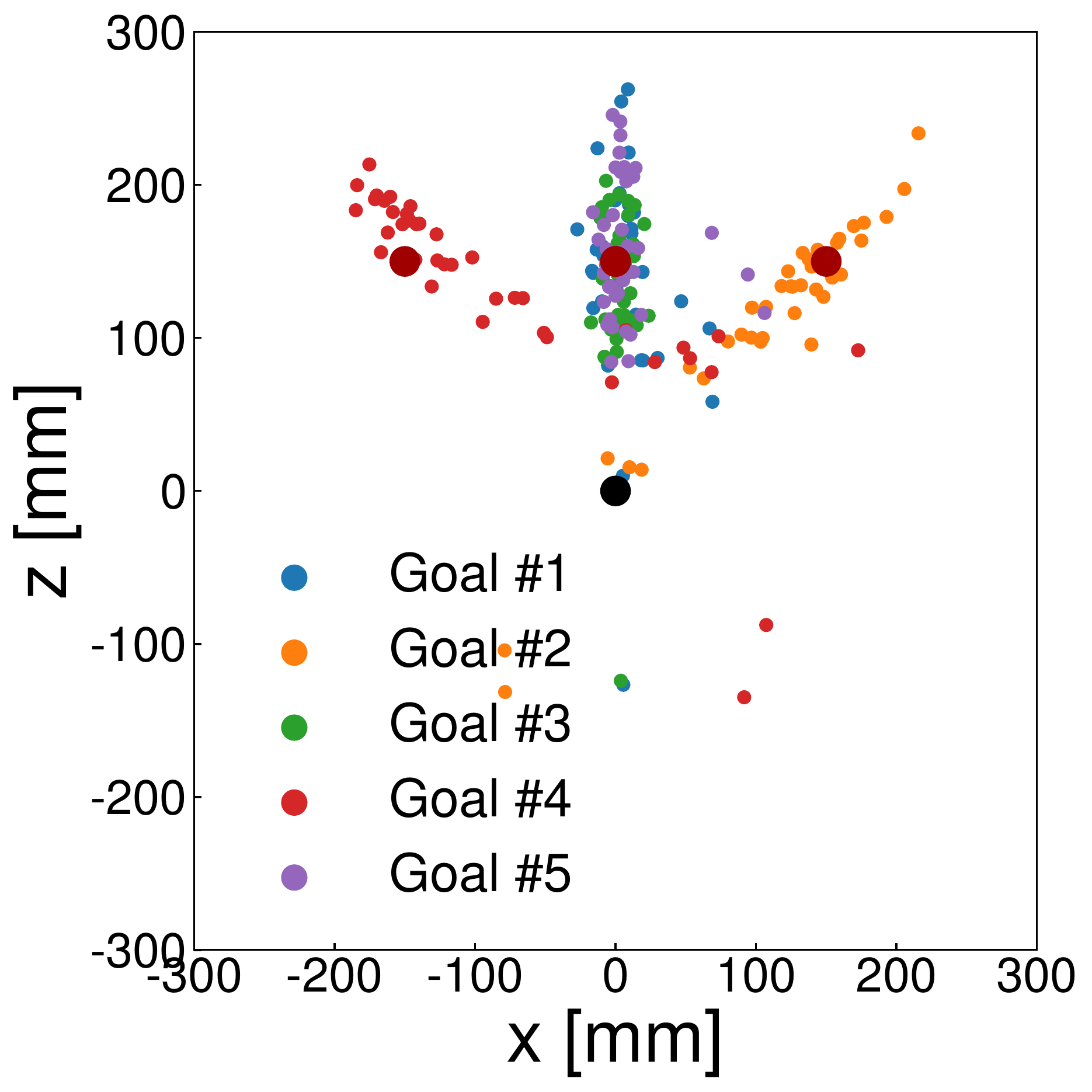}
 \end{minipage}\\
 \begin{minipage}[t]{0.5\columnwidth}
  \centering
  \includegraphics[width=\linewidth]{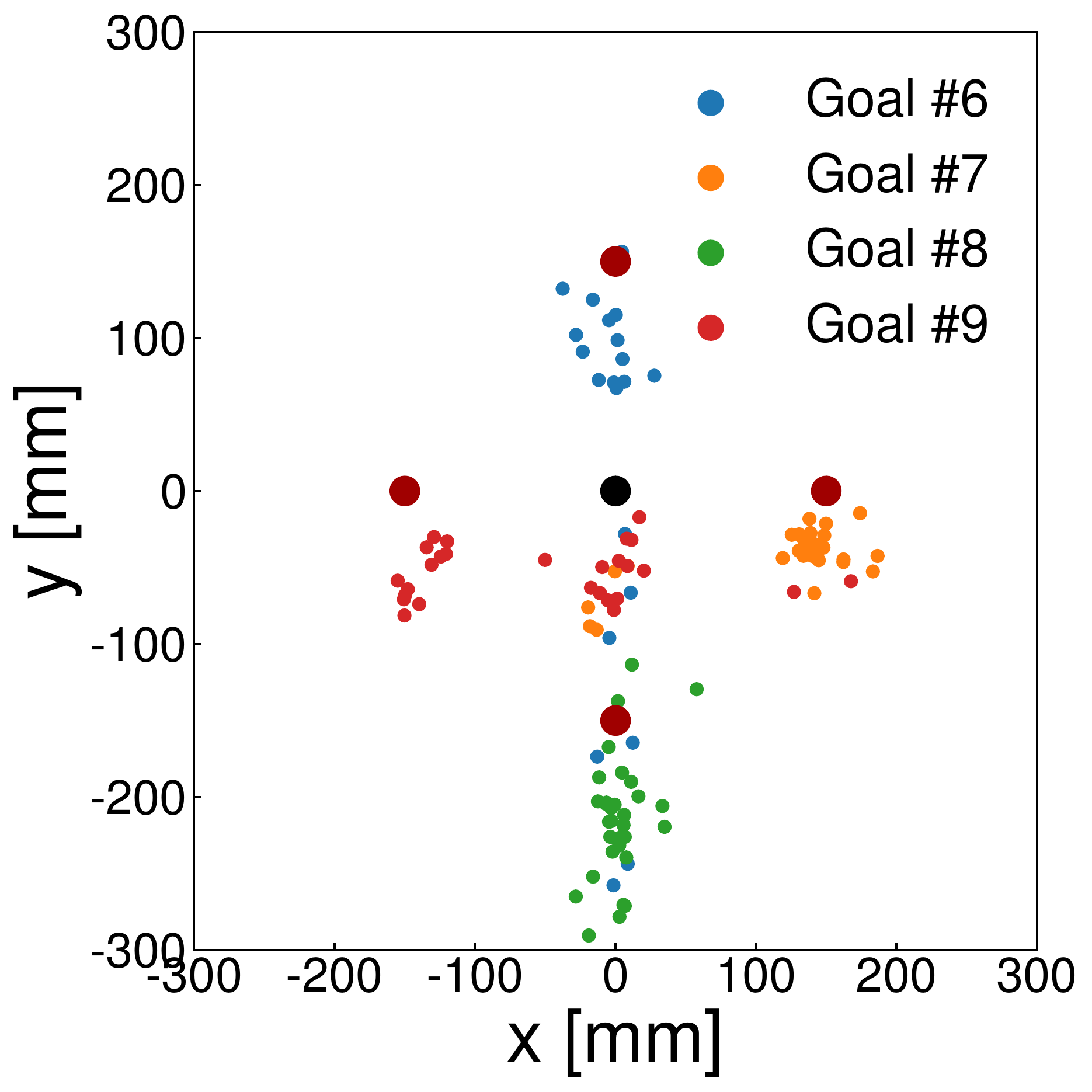}
 \end{minipage}%
  \begin{minipage}[t]{0.5\columnwidth}
  \centering
  \includegraphics[width=\linewidth]{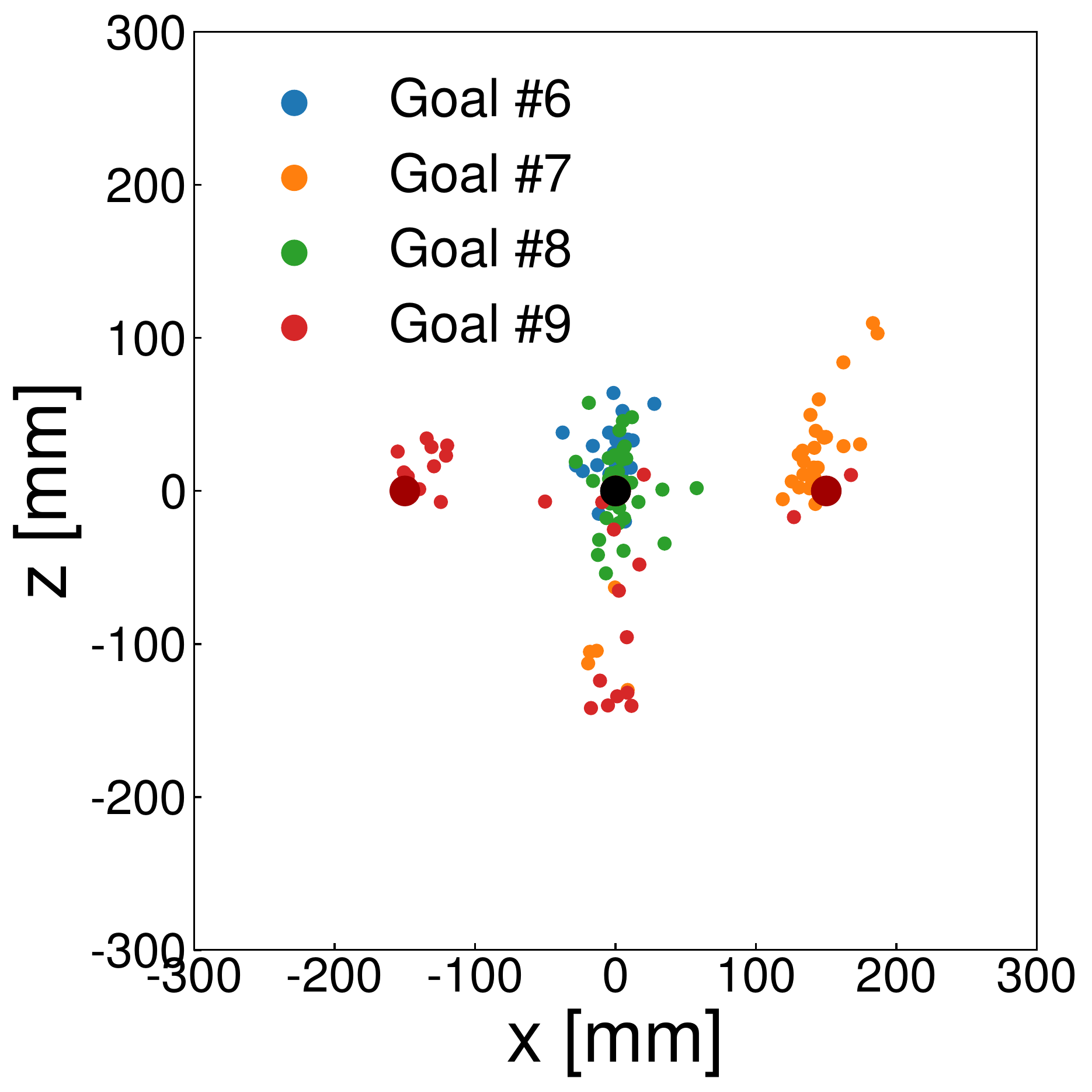}
 \end{minipage}\\
 \begin{minipage}[t]{0.5\columnwidth}
  \centering
  \includegraphics[width=\linewidth]{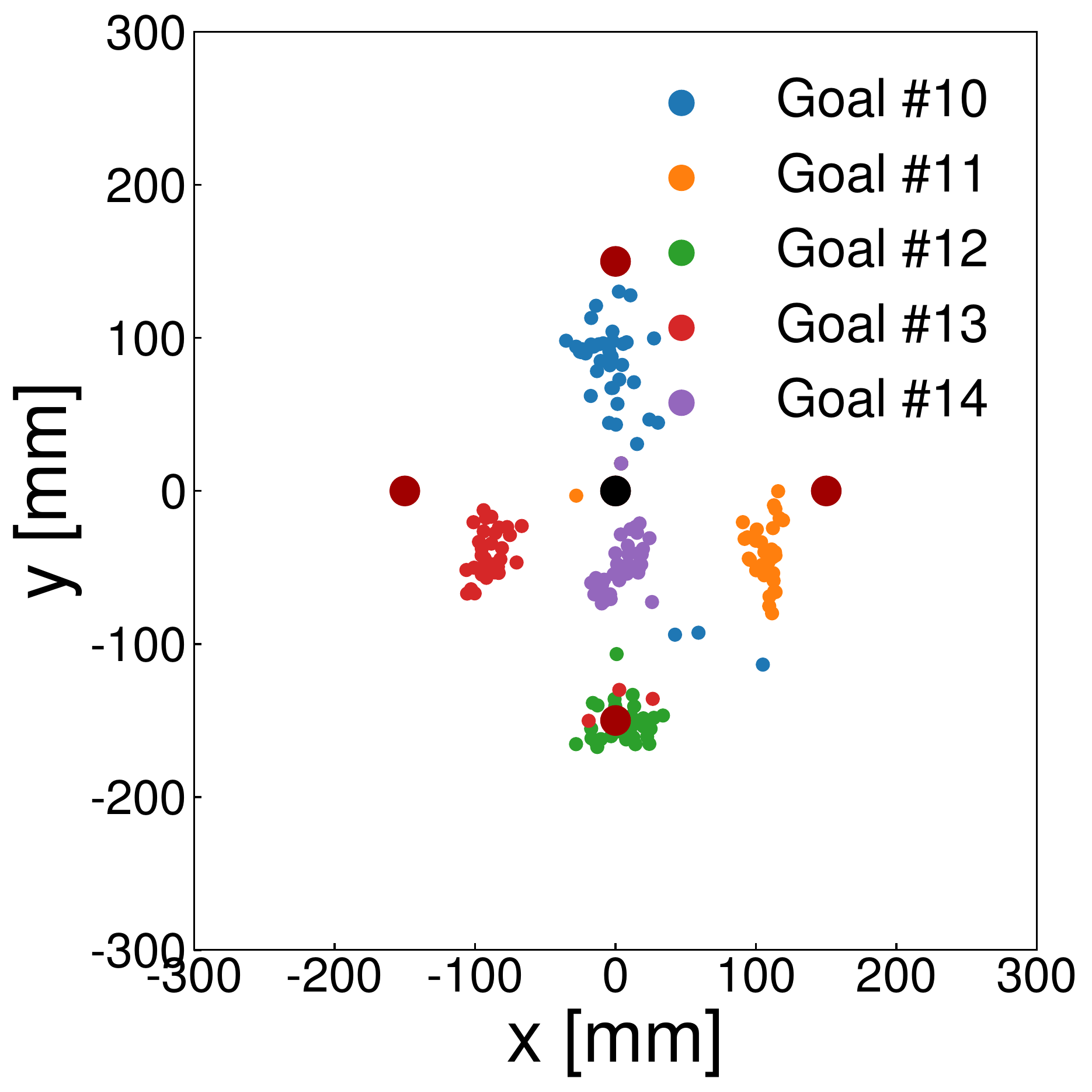}
 \end{minipage}%
  \begin{minipage}[t]{0.5\columnwidth}
  \centering
  \includegraphics[width=\linewidth]{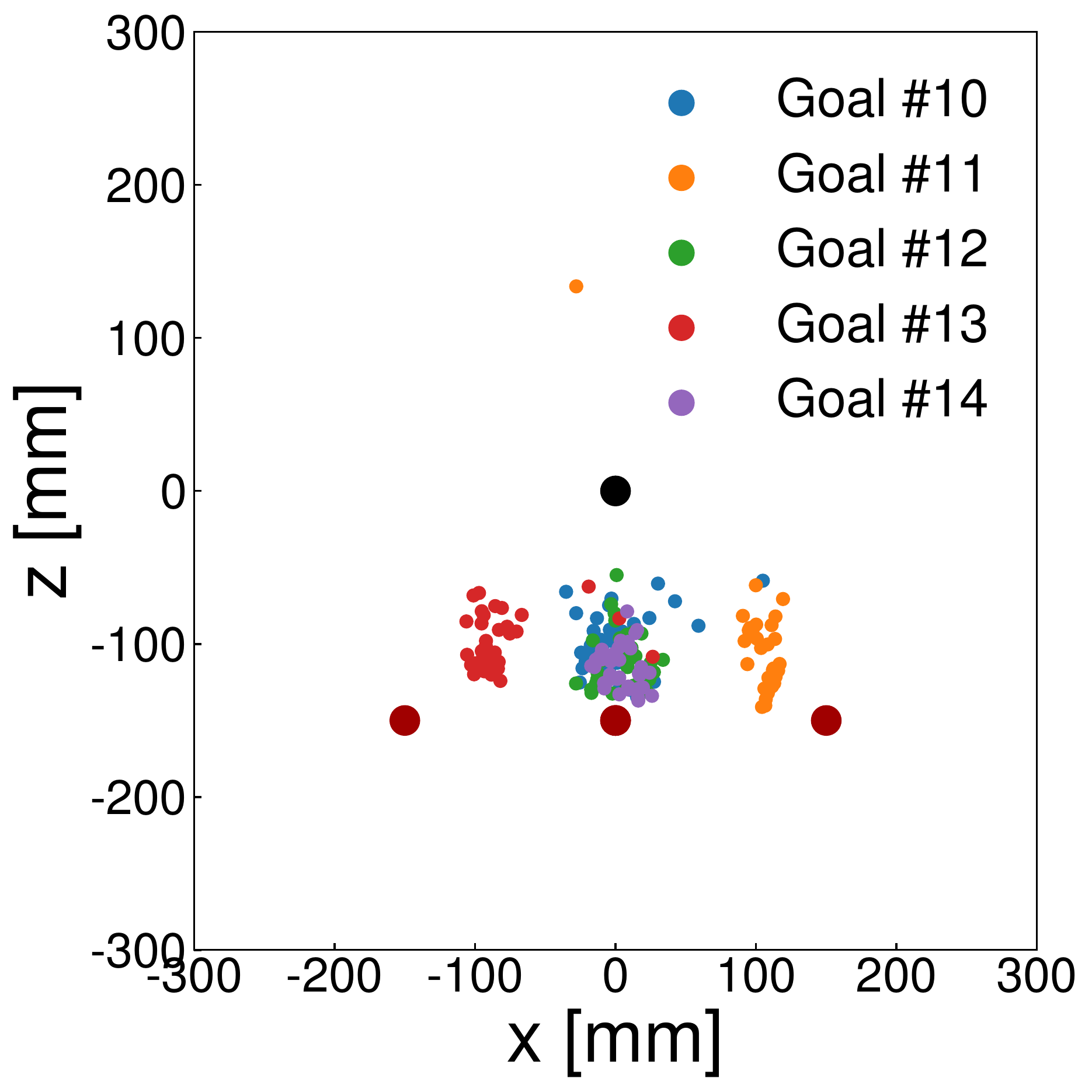}
 \end{minipage}%
 \caption{Final hand position of each trial}
 \label{fig:final_points}
\end{figure}

Fig.~\ref{fig:eps_xyz_and_time} shows Box-and-whisker plots of $\varepsilon_{xyz}$, $\varepsilon_{xy}$, and the time required to reach goal.
The median of $\varepsilon_{xyz}$ is \SI{64.34}{mm}, and the median time is 6.63 seconds.
In addition, the plot of $\bx_f$ is shown in Fig.~\ref{fig:final_points}.

Followings are participants' comments: ``A spreading circle was easier to recognize than a narrowing circle'', ``It took much time to decide the apex finally. I moved my hand many times around the apex and compared the tactile sensations of the apex and the bottom,'' and ``I could easily determine the direction of guidance.''

\begin{table}[t]
  \centering
  \caption{Trial completion rate.}
  \begin{tabular}{c|ccccccc}
   Goal \# &    1 &    2 &    3 &    4 &    5 &    6 &    7 \\\hline
   rate    & 0.92 &    1 &    1 &    1 &    1 & 0.54 & 0.79 \\\\
   Goal \# &    8 &    9 &   10 &   11 &   12 &   13 &   14 \\\hline
   rate    & 0.85 & 0.72 &    1 &    1 &    1 &    1 &    1 
   \end{tabular}
  \label{table:rate}
\end{table}
 
\subsection{Discussion}

Fig.~\ref{fig:final_points} shows that the participants could move their hands in the direction of the apex, as the plots are concentrated around the goal or on the line from the start to the goal.
The median error $\varepsilon_{xy}$ was \SI{40.41}{mm}, indicating that the participants felt the circle within the palm area. 

The median $\varepsilon_{xyz}$ was \SI{64.34}{mm}. 
The error came from the judgment error of the cone apex.
In this experiment, the minimum diameter of the circle is set to be \SI{1}{cm}.
Thus, the size of the circle does not change within \SI{25}{mm} around the goal in the case of vertical and horizontal guidance, and within $\SI{25}{mm}\times\sqrt{2}\sim\SI{35}{mm}$ in the case of diagonal guidance.
Therefore, this degree of error is inevitable.
In addition, the deviation of \SI{64}{mm} from the apex corresponds to the radius difference of \SI{7.8}{mm} in cross-section, which was difficult to distinguish for the participants.
Therefore, we think that the error $\varepsilon_{xyz} = \SI{64.34}{mm}$ is reasonable enough.

Furthermore, this is also considered due to an error between the sensing position and the area where the participants perceived the sense of touch.
In the present experiment, the system is designed to sense the center coordinates of the palm.
However, it is unclear which part of the palm the participant perceived the tactile sensation.

As shown in Fig.~\ref{fig:final_points}, there were some trials in which the hand was moved opposite the guiding direction and finished the trial.
The circle presented in this paper is designed to become larger as one moves away from the apex.
In this experiment, the cross-section circle was presented by discrete ten points.
Therefore, if the subject continues to move their hand in the direction opposite to the direction of the hand guidance, the gap between points becomes larger than the size of the palm at a certain moment, and only one point is presented on the palm.
Therefore, the participants sometimes misrecognized this one point as the apex.
This problem can be solved by increasing the number of sampling points or sampling only the area where the cone and the hand overlap.

Table.~\ref{table:rate} shows that the trial completion rate of the horizontal guidance (Goal No. 6 to 9) is lower than that of the upward and downward guidance.
That means that the participants sometimes lost the tactile stimuli and completely lost the clue to the direction.
In the case of upward/downward guidance, tactile sensations are always presented to the hand according to its height.
In the horizontal case, however, no tactile sensation is presented when the hand is displaced in the height direction. Therefore, it is considered that the trial could not be terminated because the tactile sense to recognize the guiding direction was lost. 

In this experiment, the displayed cone was fixed in space.
We consider it possible to guide the hand in a long distance more stably by improving the displayed dynamic cone.
When the user’s hand goes out of the cone, the hand is guided by a newly created cone to the goal.
When the guiding distance is too long, the hand should be guided by multiple cones being switched according to the hand position. 
Switching the cone periodically keeps the apex angle large, making it easy to perceive the motion direction.

We also think the time to reach the goal point will be shorter by improving the sensor feedback.
In the current system, the frame rate of the sensor is \SI{30}{fps}, and there is a delay of about \SI{90}{ms} from sensing to the presentation of ultrasonic waves.
Therefore, the hand movement speed was limited to less than \SI{0.45}{m/s}, and feedback could not follow if they exceeded the speed.

\section{Conclusions}

In this study, we proposed a method to guide a hand to a target positon in space.
The user moves their hand toward the apex of a haptically created cone, perceiving the cross-section of the cone produced by ultrasonic tactile stimuli.
The performance was evaluated by participant experiments.
As a result, it was found that the system can guide a hand to an arbitrary target positon in the three-dimensional space without visual and auditory information.
In particular, it was shown that in the case of downward guidance, approaching the phased array, the guidance was more accurate than the upward or horizontal direction.
As a result, the median error of the goal and hand position was \SI{64.34}{mm}, and the median time to complete the trial was 6.63 seconds in a \SI{30}{cm} cube workspace, except for those who could not complete the trial.
These results can be improved by dynamically switching the cone and reducing the delay of the sensor feedback.
As future work, we would like to address the problem of identifying the optimum conditions for guidance and the dynamic switching of cones.

\section*{ACKNOWLEDGMENT}
% xxxxxxxxxxxxxxxxxxxxxxxxxxxxxxxxxxxxxxxxxx\\xxxxxxxxxxxxx

\bibliography{ref}
\bibliographystyle{IEEEtran}

\end{document}